\documentclass[twocolumn,showpacs,preprintnumbers,amsmath,amssymb]{revtex4}

\usepackage{graphicx,color}
\usepackage{dcolumn}
\usepackage{bm}
\usepackage{subfigure}

\usepackage{epstopdf}

\begin{document}


\title{Size Dependent Growth in Metabolic Networks}

\author{Henry Dorrian$^1$\footnote{H.Dorrian@mmu.ac.uk},
Kieran Smallbone$^2$\footnote{Kieran.Smallbone@Manchester.ac.uk} and
Jon Borresen$^1$\footnote{J.Borresen@mmu.ac.uk}}

\affiliation{ $^1$ Dept of Computing and Mathematics, Manchester
Metropolitan
University, Manchester UK M1 5GD\\
$^2$ School of Mathematics, The University of Manchester,\\
Manchester UK M60 1QD}

\date{\today}

\begin{abstract}

\noindent Accurately determining and classifying the structure of complex
networks is the focus of much current research. One class of network
of particular interest are metabolic pathways, which have previously
been studied from a graph theoretical viewpoint in a number of ways.
Metabolic networks describe the chemical reactions within cells and
are thus of prime importance from a biological perspective.

Here we analyse metabolic networks from a section of microorganisms,
using a range of metrics and attempt to address anomalies between
the observed metrics and current descriptions of the graphical
structure. We propose that the growth of the network may in some way
be regulated by network size and attempt to reproduce networks with
similar metrics to the metabolic pathways using a generative
approach.  We provide some hypotheses as to why biological networks
may evolve according to these model criteria.
\end{abstract}

 \maketitle

The graphical structure of metabolic pathways has been extensively
studied
\cite{Lacroix2008,jeong2000large,wagner2001small,schuster2011topological}
and describing the structure should give insight into functionality
\cite{Stelling}. Many of the salient features of such networks have
been investigated, particularly any scale free nature
\cite{barabasi2000scale} although the scale free model is currently
the subject of some debate \cite{Khanin,Stumpf}. Of particular
interest are the high clustering coefficients observed in metabolic
pathways which have previously been explained using concepts of
topological hierarchy \cite{ravasz2003hierarchical} and modularity
\cite{Takemoto}. Other concepts describing the structure such as
memory \cite{Klemm} and decomposition into functional modules
\cite{Ma} have also been proposed.

A more recent model by Schneider et al. \cite{schneider2011scale}
uses network depletion, where a fully connected network is degraded
according to the degree of the nodes.  This gives rise to the high
clustering coefficient and lower degree distribution values found in
metabolic pathways, but the model cannot agree with how these
biological networks would have evolved (i.e  it is not plausible
that metabolic networks have evolved from being fully connected and
slowly losing connections as, for one thing, not all metabolites can
react with one another).

Previous work by Dorogovtsev and Mendes \cite{Dorogovtsev} explains the presence of high
degree decay rates in evolving networks with accelerating growth
rates, where it is shown that for networks where the growth is
accelerating and where the decay rate of the degree distribution
$\gamma>2$, the probability distribution for preferential attachment
must be non-stationary.

It has also been shown that the probability of a
metabolite reacting with k other metabolites decays as $P(k) \sim
k^{-2.2}$ \cite{jeong2000large,wagner2001small}. However the
metabolic pathways investigated here demonstrate $\gamma<2.2$ in all cases.  This
leads to the biologically plausible idea that the growth of such
networks may be slowing and that the probability of connection to
existing nodes may not be static.

Initially, we conduct a graphical analysis of eight bacterial
metabolic pathways, concentrating on their clustering coefficients
and mean path lengths. Here microorganisms have been chosen, as the
metabolic pathways are observed to be less modular than higher
organisms allowing greater illustration of the concept. Further to
this, a growth model, whereby the rate of growth decays as a
function of network size, is used to demonstrate that size dependent
growth may provide a suitable explanation as to many of the
structural features of metabolic networks.

\section{Graphical Analysis}

Eight microbial metabolic pathways were considered, they were as
follows:

\noindent \textit{Escherichia Coli, Escherichia Coli iAF1260,
Escherichia Coli iJR904,
 Helicobacter Pylori, Methanosarcina Barkeri, Staphylococcus Aureus,
 Mycobacterium Tuberculosis and Saccharomyces Cerevisiae.}\\

Three formulations of E Coli are chosen to ensure the results are
independent of the methodology used to initially determine the
network.

The eight metabolic reconstructions are downloaded in SBML
format~\cite{SBMLonline} from the BiGG database
\cite{schellenberger2010bigg}. The models were imported to Matlab
using libSBML~\cite{libSBML}.

We consider each metabolic pathway as an undirected graph with the
adjacency matrix of the graph being the boolean representation of
the chemical interactions. For each model, nodes $i$ and $j$ are
defined as adjacent if metabolite $i$ appears as a reactant and $j$
as a product, or indeed $i$ as a product and $j$ as a reactant in
any reaction.  Although this is a highly simplified representation
this has been shown to be a useful tool in analysing such systems
\cite{schuster2000general}.

Note: There are a variety of network constructions available, both
including and excluding subcellular compartmentalisation, and
with/without considering the role of water and protons.  We consider
a formulation with subcellular compartmentalisation, as this most
accurately represents the structural nature of the biochemical
processes within the cells and excluding the presence of water and
protons as is customary in such studies as these
\cite{Lacroix2008,jeong2000large,wagner2001small,schuster2011topological}.
The results obtained here are applicable to the other available
formulations with some small modification to the size dependent
decay constant $C$ described in Section ~\ref{sec:SDGM}.

\subsubsection{Clustering Coefficients}

The clustering coefficient of a network is a measure of transitivity
- how the nodes in the network tend to cluster together. We consider
a network average global clustering coefficient
\cite{watts1998collective} of the form.

\begin{equation}
<c>=\frac{1}{n}\sum_{i=1}^n c_i,
\end{equation}

\noindent where $n$ is the number of nodes in the network,

\begin{equation}
c_i=\frac{2e_i}{k_i(k_i-1)},
\end{equation}

\noindent where $k_i$ is the degree of node $i$ and $e_n$ is the number of connected
pairs between the nodes to which node $i$ is connected.

The observed clustering coefficients $<c>$ for the metabolic data
show that the networks are highly clustered and that the clustering
coefficient is independent of network size (see figure
\ref{fig:MetsAndFits}).

\subsubsection{Mean Minimum Path Lengths}

The minimum path length (or geodesic length) is a measure of the
smallest number of nodes between any two nodes, and represents the
shortest route along the network between them. For our metabolic
pathways, the average path length is surprisingly low given both the
size of the networks and their average degree. This demonstrates a
very strong small world effect \cite{milgram1967small}. Conversely,
the highest of the minimum path lengths are somewhat greater than
one would expect, given such a small world effect. For instance, the
H. pylori metabolic network with $562$ nodes has an average path
length of $2.8$ and a maximum path length of $8$. For a non-directed
network of this size, such a high maximum path length suggests
structural qualities not in keeping with other small world networks.

\subsubsection{Scale Free Structure}

It has been previously observed that the graphical structures of
metabolic pathways have much in common with many other complex
networks, particularly with respects to their possible `scale-free'
nature \cite{jeong2000large,fell2000small}.

In a scale-free network, the probability, \(P(k)\), of a node in the
system having \(k\) connections follows a power law distribution of
the form \(P(k)\sim k^{-\gamma}\) \cite{barab1999emergence}. This is
in contrast to the much studied random graphs which follow a Poisson
or Binomial distribution \cite{erds1961evolution} but are similar to
the social networks Milgram \cite{milgram1967small} described.

As can be observed (see figure \ref{fig:SAureusSD}), the metabolic
networks display some scale-free property, however they cannot be
considered truly scale free as there are too few connections with
low degree and too many with high degree for an accurate fit of the
form \(P(k)\sim k^{-\gamma}\) to be valid. Approximations for the
value $\gamma$ (above) for the metabolic pathways via least squares
fits, obtain values in all cases of $\gamma <2$ - although such a
logarithmic fit of the data yields high least square errors.

Networks which obey a power law distribution of the form \(P(k)\sim
k^{-\gamma}\) can be artificially generated. For instance
Barab{\'a}si and Albert \cite{barab1999emergence} describe a network
growing via {\em{preferential attachment}} of new nodes to existing
nodes with a higher degree, the probability that a new node connects
to an existing
node is calculated using equation ~\ref{eqn:SF}.\\

\begin{equation}
P(\textit{New node connects to node i})=\frac{K_i}{\sum_j (K_j)}
\label{eqn:SF}
\end{equation}

\noindent where $K_i$ is the degree of node $i$ and the sum is over all
pre-existing nodes $j$.

Artificially generated networks of this type typically have $\gamma
\in [2, 3]$ - greater than the observed metabolic data.  The
clustering coefficients are also considerably lower than those
observed for the metabolic networks (typically lower than $0.1$ for
any randomly generated Barab\'{a}si– Albert model (BA model)) and
are observed to decrease with increasing network size.

\subsubsection{A Size Dependent Generative Model} \label{sec:SDGM}

Although the BA model (equation \ref{eqn:SF}) does not provide good
fits to the metabolic networks, a generative model using
preferential attachment would seem to have much to offer when
considering the growth of metabolic networks. Here promiscuous
metabolites within the network with high degree ie those which are
present as reactants in higher numbers of reactions will be those
which are more likely to form new connections, whereas co-factors
which are more specific in their function will form less
connections.

As such, a generative model similar to the BA model appears as a
strong candidate for describing some of the features which develop
in the growth of metabolic networks. However, as a variety of
studies have demonstrated
\cite{schneider2011scale,ravasz2003hierarchical,ravasz2002hierarchical},
the scale-free model alone is not sufficient to fully describe the
networks when a range of graph metrics are considered.

In considering a generative model approach to recreating networks
similar to the metabolic networks of microorganisms, the concept of
size dependent growth was considered.  As such we attempt to
introduce some concept of limiting factors on the generative model.
Essentially we attempt to model the growth of the network such that,
initially, it is very easy for new connections to form and as the
network grows the probability of new metabolites attaching is
reduced. We have chosen a linear model for simplicity.

Networks are grown according to the following:

\begin{equation}
P(\textit{New node connects to node i})=\frac{K_i}{\sum_j (K_j)}
\times \frac{C}{n} \label{eqn:SDGM}
\end{equation}

\noindent where $C$ is some constant and $n$ is the number of nodes in the
existing network. Such a model will produce a globally connected
network for $n^2+1 \leq C$. For $n^2+1 > C$ the preferential
attachment model begins, however, the probability of attachment is
initially high.  As the network grows and $\sum j \sim C$, new nodes
attach in a manner identical to the original BA model and for $\sum
j > C$ the network ceases to grow any further.  (Note, this model
assumes that nodes are not self connected and restricts the
probability of any new attachment to $P \leq 1$, for $n^2+1 \leq C$).

The effect of modifying the BA model thusly has two effects: Firstly
the probabilities of attachment are not static and are rescaled as
each new node enters the network, secondly the probability that any
new node will attach is decreasing as the network grows.

\begin{figure}[!Ht]
\begin{center}
        \includegraphics[width=3in, height=1.5in]{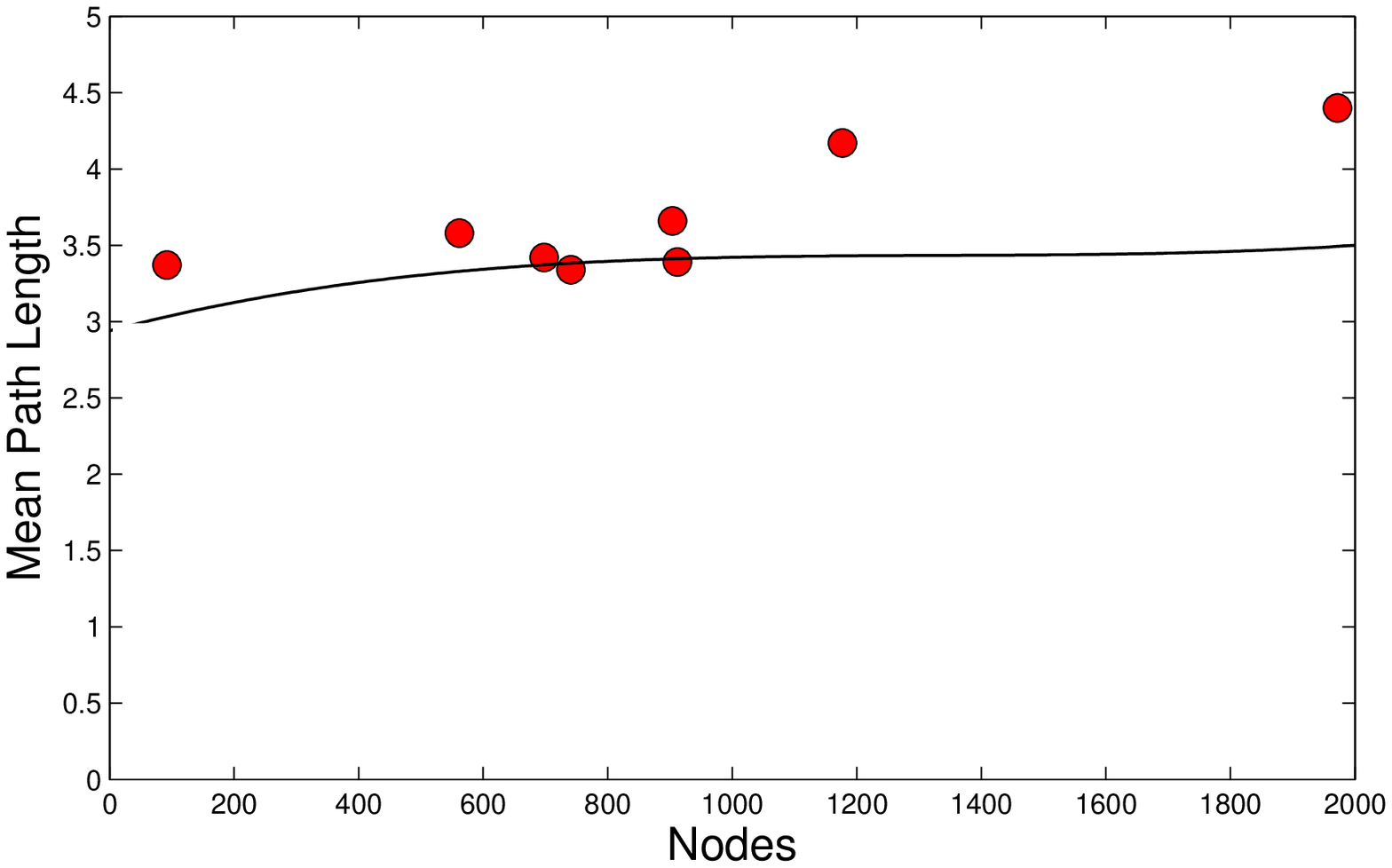}
        \includegraphics[width=3in, height=1.5in]{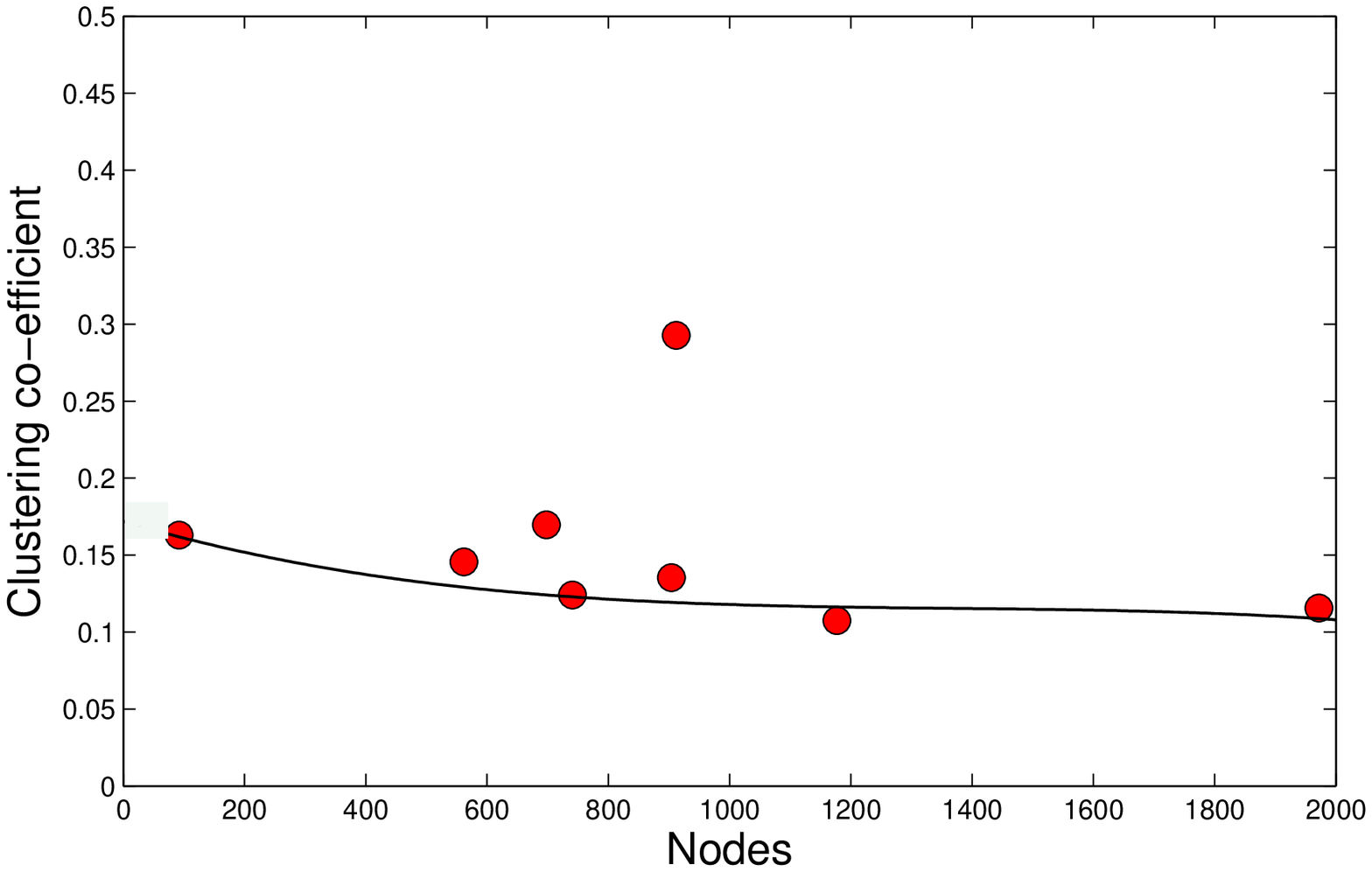}
        \vskip 0.1cm
\caption{Mean minimum path lengths (top) and clustering coefficients(bottom) for
eight microbial metabolic pathways (red circles) and (black line)
average minimum path lengths and clustering coefficients for
randomly generated size dependent networks with $C=N/2$ (equation
\ref{eqn:SDGM}). } \label{fig:MetsAndFits}
\end{center}
\end{figure}

The size to which any network will grow in finite time, is
essentially determined by the value $C$.  When $n>C$, the
probability of new nodes attaching is very small and these are new
nodes are rejected.  Simulations were conducted until the network
had grown to a specified size, attempting to attach new nodes until
this was successful.

Repeated simulations for network growth according to model
\ref{eqn:SDGM} were performed for varying constant $C$.  It was
observed that for $C=N/2$, where $N$ is the size to which the
network is grown, the average path length and the clustering
coefficients of the generated networks fitted the metabolic pathway
data better than any single previous structural description of
metabolic networks (see figure \ref{fig:MetsAndFits}).

It should be noted that the networks generated are not truly
scale-free, being somewhere a hybrid of both exponential and scale
free type distributions which is not unlike the original metabolic
networks.

The overall effect of such a size dependent modification to the
original BA model is to initially produce a highly connected 'hub'
which then grows via preferential attachment giving rise to high
clustering coefficients and a very strong small world effect
resulting in low average path lengths.

It is straightforward to amend the size dependent model
\ref{eqn:SDGM} such that no globally connected initial stage occurs
by rescaling the size dependent modification in Equation \ref{eqn:SDGM}
to $C/ n+\sqrt{C}$. This produces a network which has a more
pronounced scale free structure, while retaining the high clustering
coefficients of the original formulation.

\subsubsection{Example - S Aureus}

The bacteria Staphylococcus Aureus has a metabolic network of $741$
nodes, a clustering coefficient of $<c>= 0.124$, a maximum geodesic
length of $9$ and a mean geodesic length of $3.34$.

The (mean) average values for $100$ trials of a network of $741$
nodes, generated according to equation \ref{eqn:SDGM} with $C=37.5$
are: clustering coefficient $<c>=0.125$, maximum path length $7.84$
and mean geodesic length $3.53$.

\begin{figure*}[!ht]
\begin{center}
        \includegraphics[width=2in, height=1.5in]{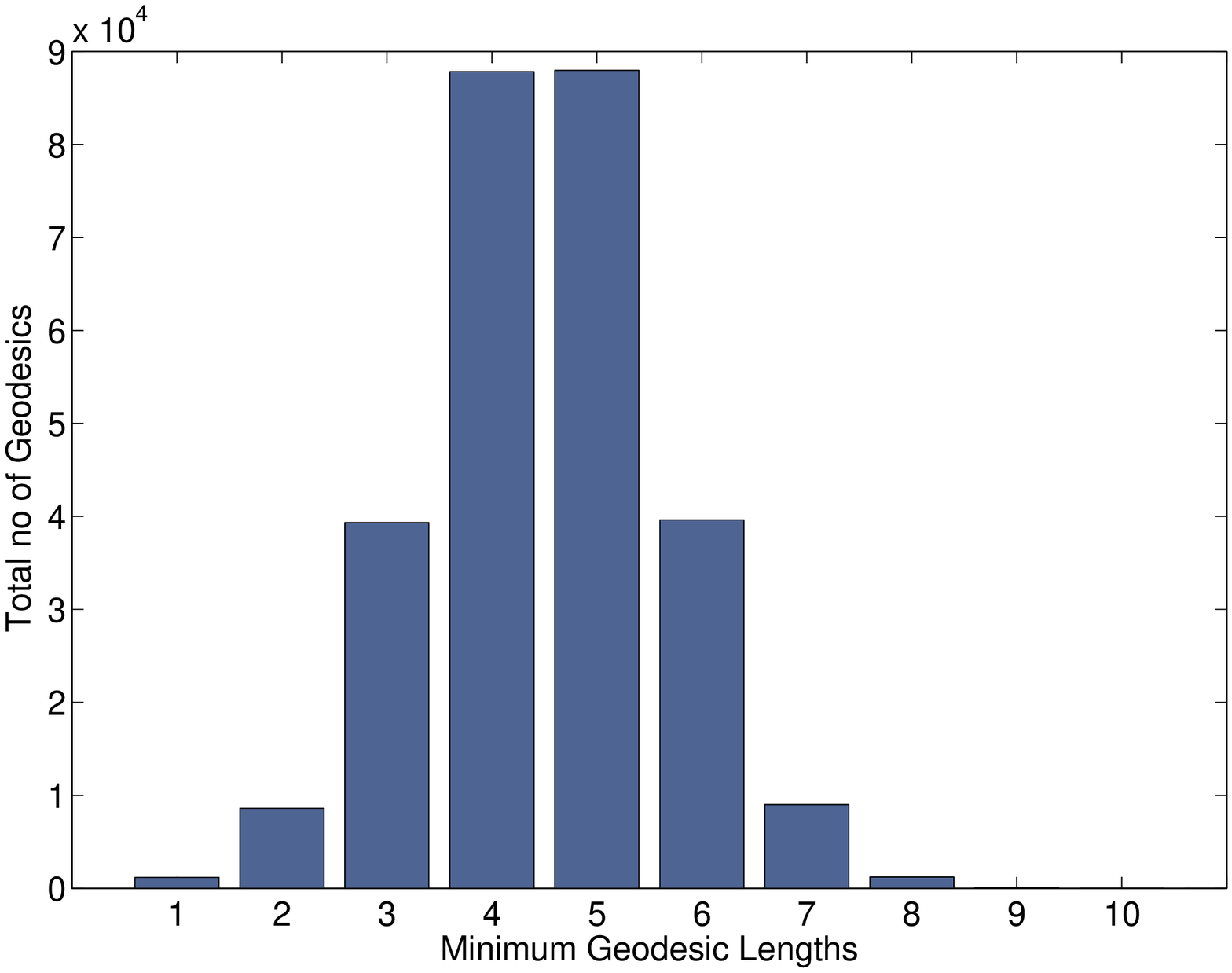}~~
        \includegraphics[width=2in, height=1.5in]{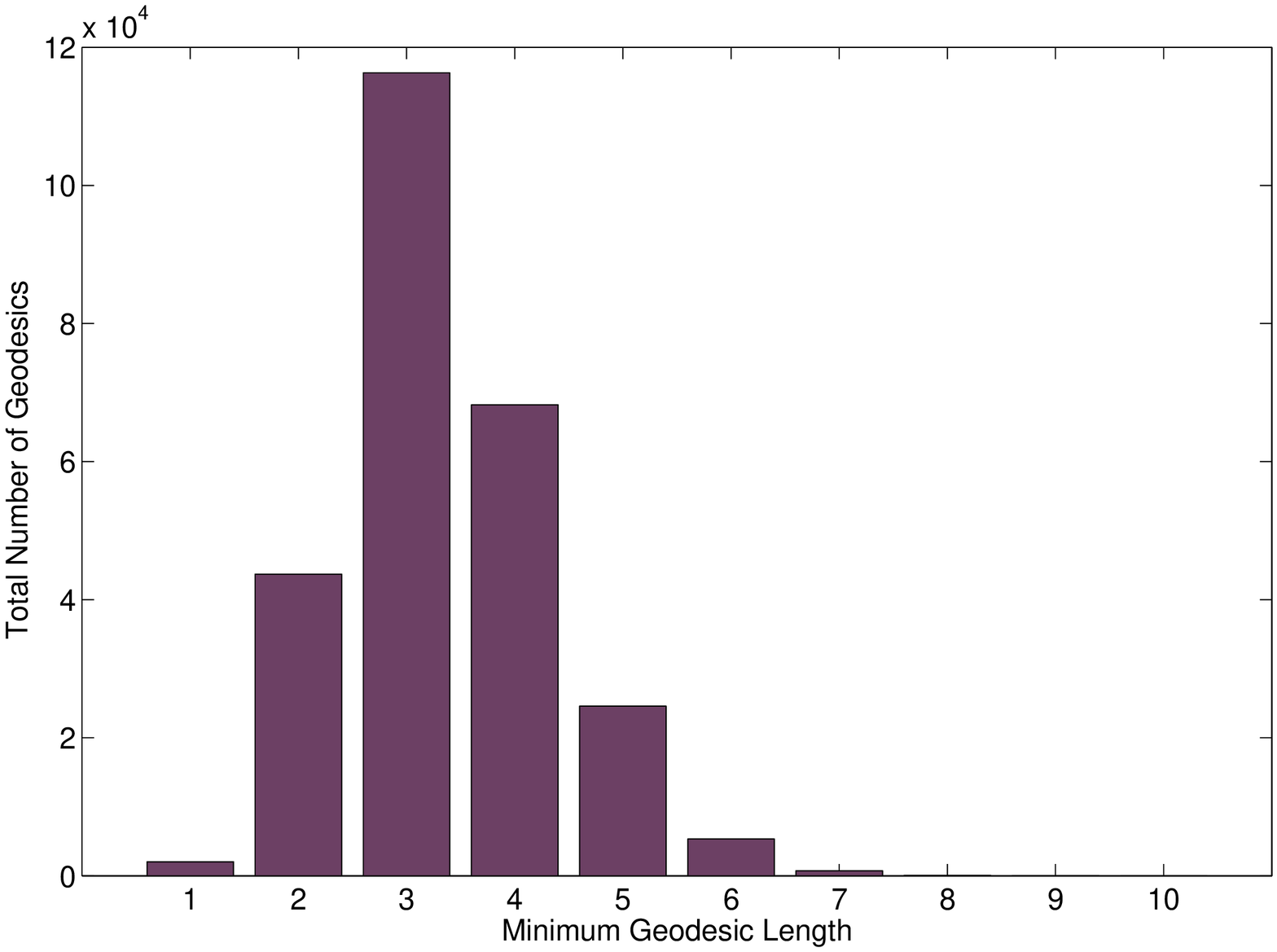}~~
        \includegraphics[width=2in, height=1.5in]{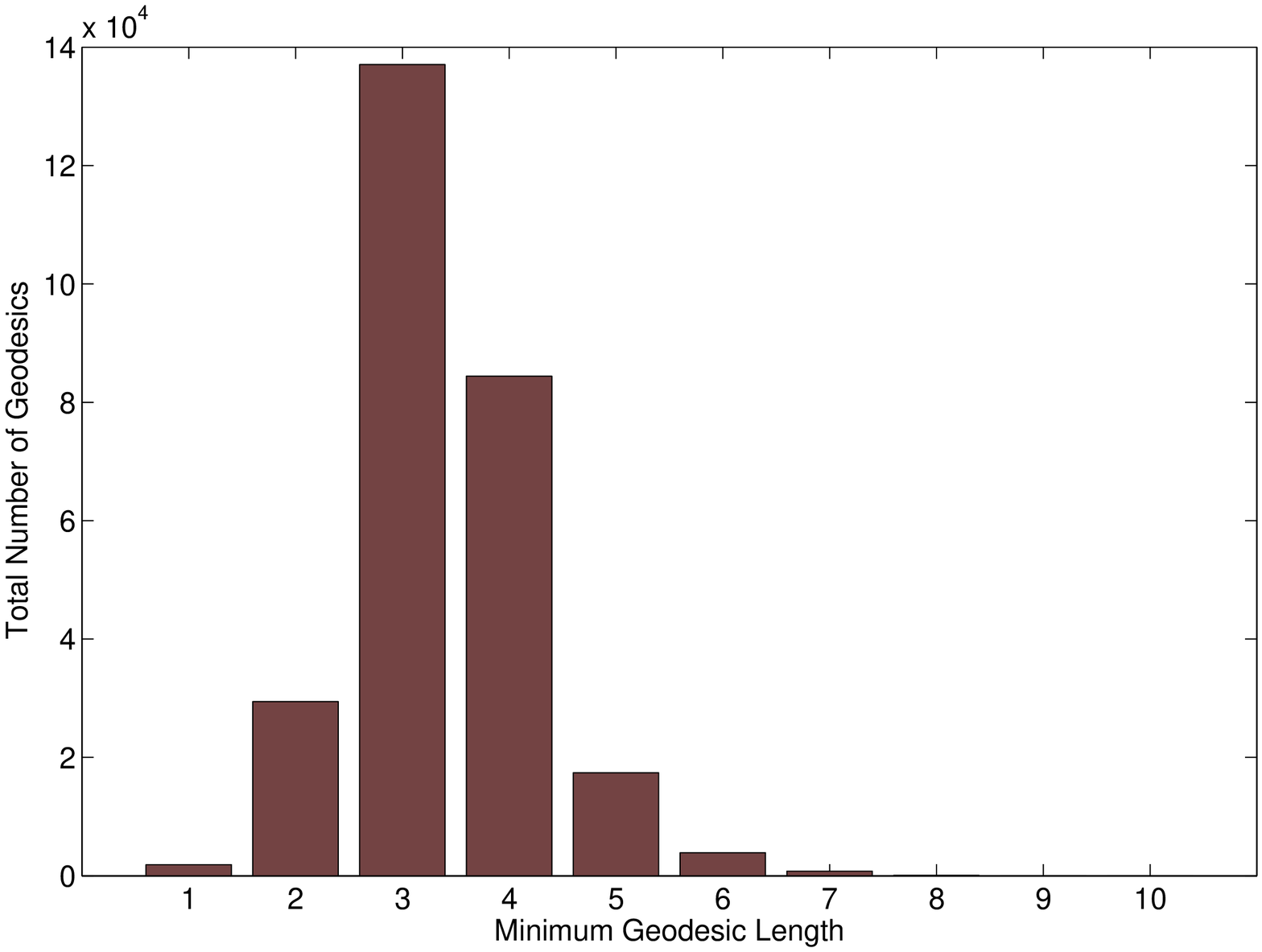}
        \includegraphics[width=2in, height=1.5in]{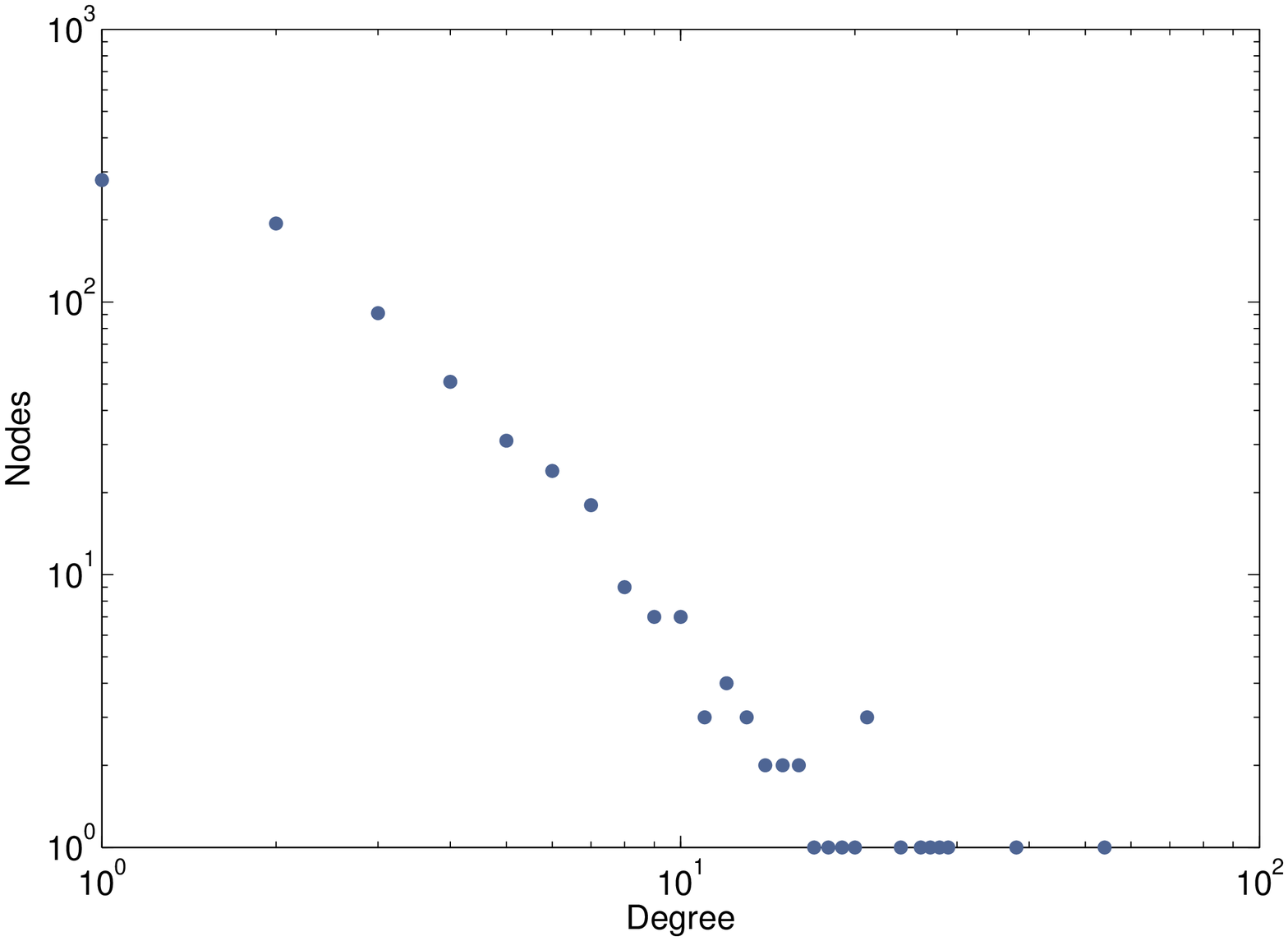}~~
        \includegraphics[width=2in, height=1.5in]{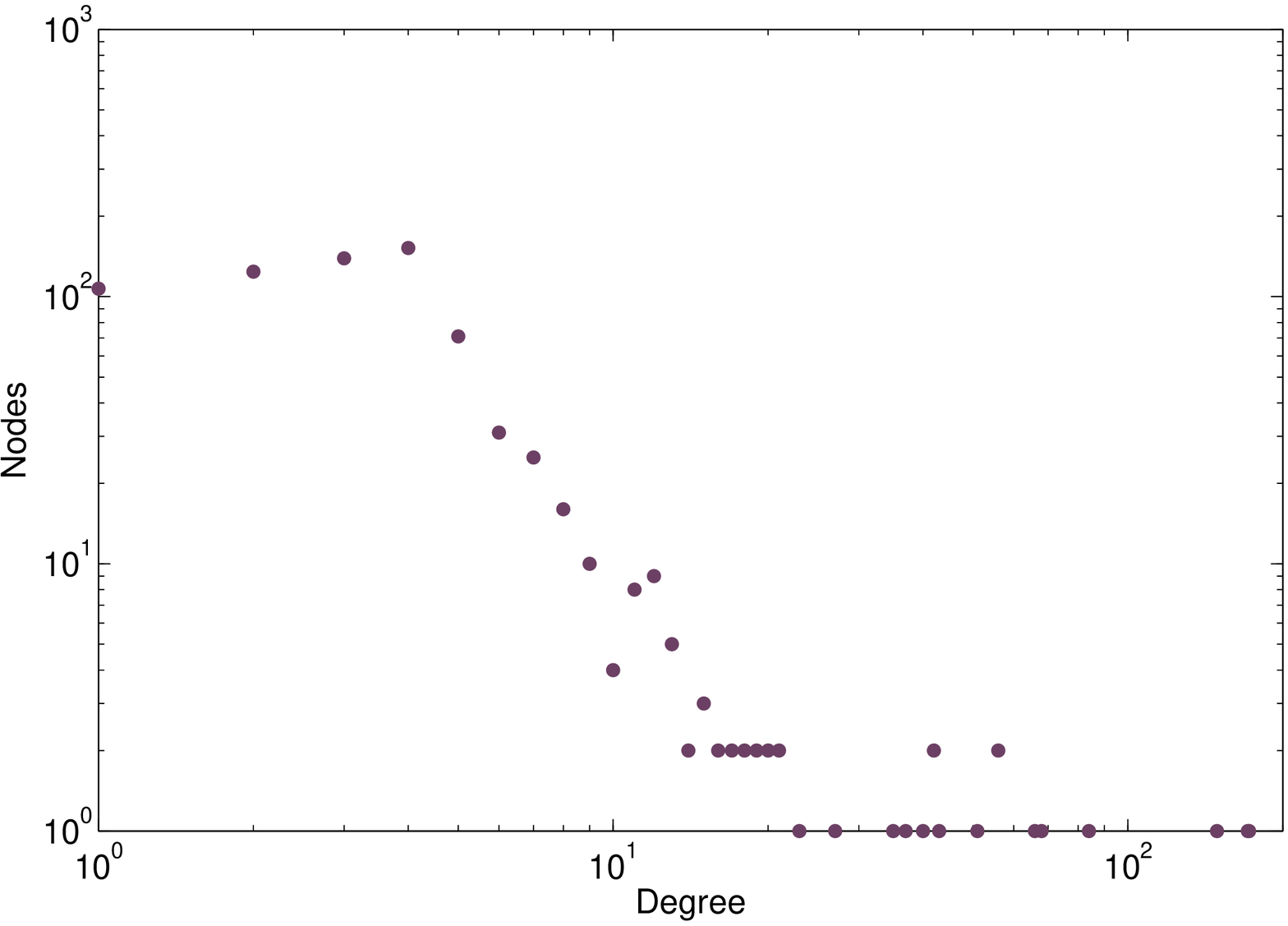}~~
        \includegraphics[width=2in, height=1.5in]{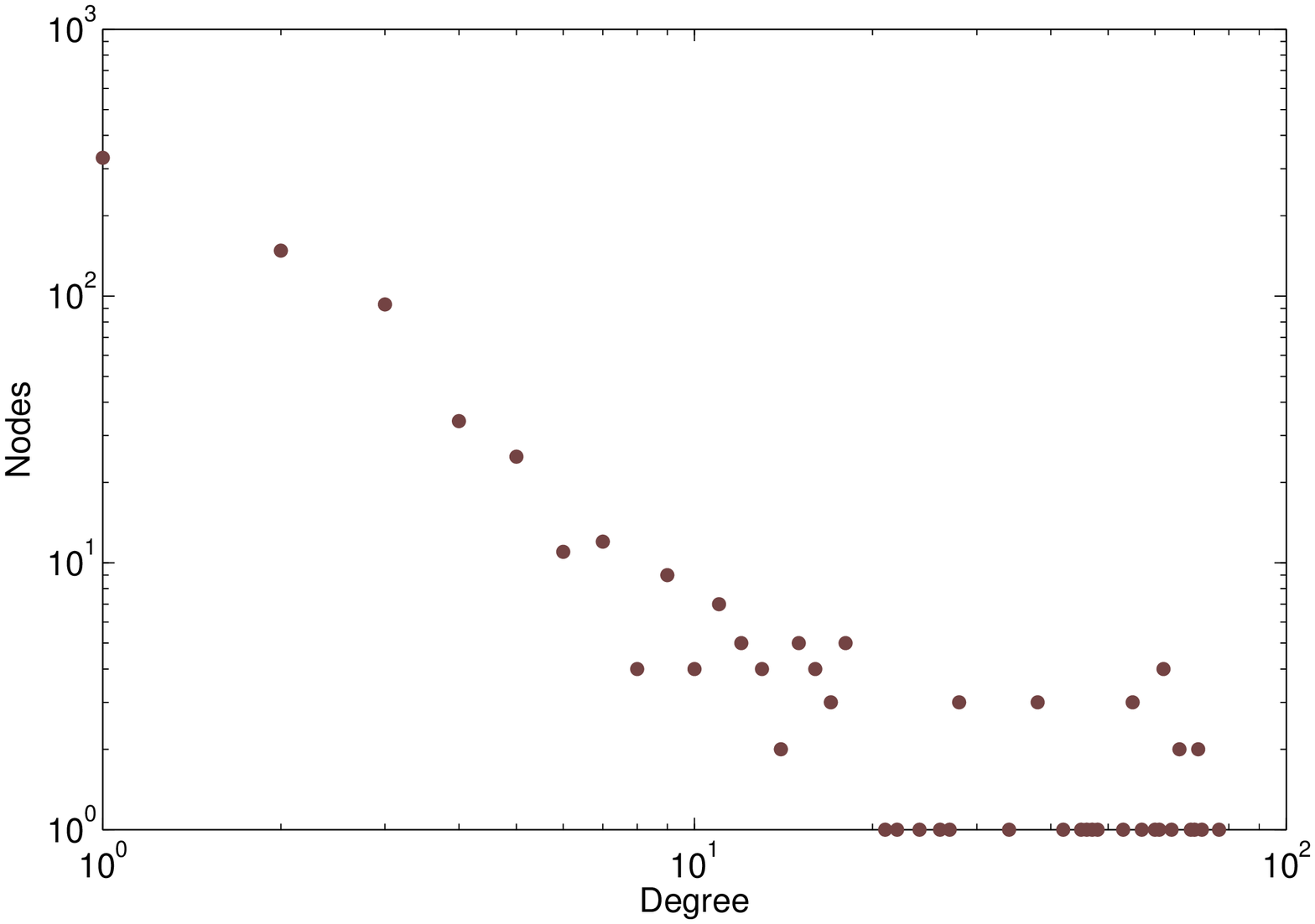}
\caption{Minimum geodesic lengths (top) and degree distribution
plots (bottom) for S Aureus and example size dependent network of
$741$ nodes.  Note the geodesic distribution for the size dependent
network matches closely the metabolic network.  The degree
distribution for the metabolic network is not entirely scale free
due to fewer nodes of degree $1$ and $2$ than would be expected. The
size dependent growth model demonstrates a degree distribution which
is in many respects closer that of the metabolic pathways.
\ref{eqn:SF}} \label{fig:SAureusSD}
\end{center}
\end{figure*}

The distribution of geodesic lengths over the whole of the S Aureus
network and an example size dependent network are shown in figure
\ref{fig:SAureusSD} in addition to the degree distributions.

\begin{figure*}[!ht]
\begin{center}
    \includegraphics[width=2in, height=2in]{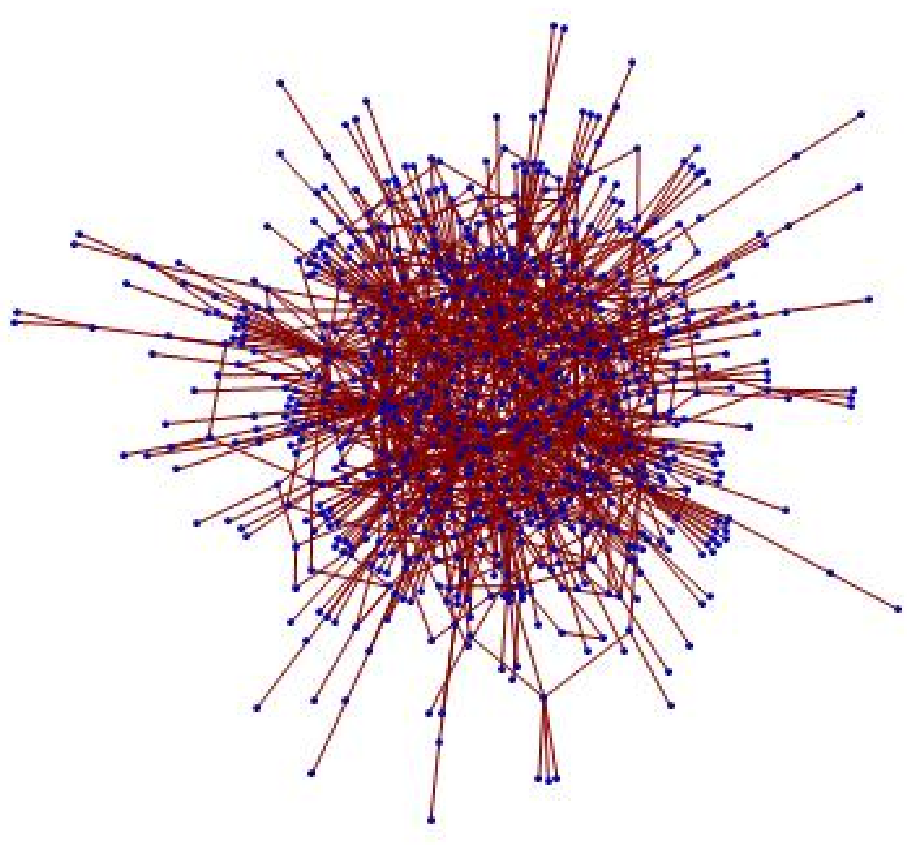}
    \includegraphics[width=2in, height=2in]{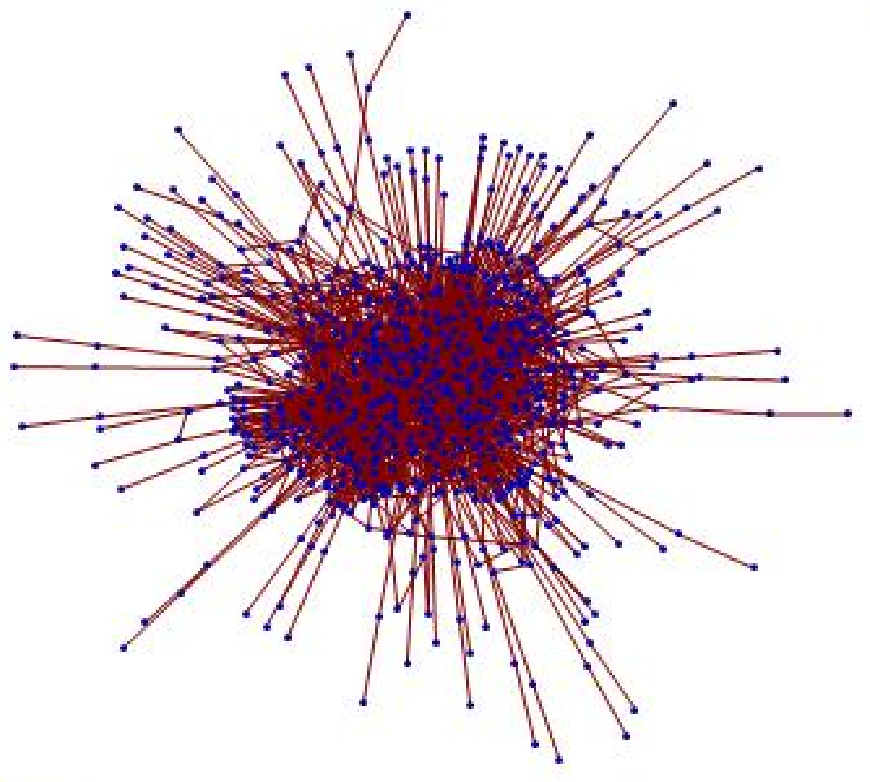}
    \includegraphics[width=2in, height=2in]{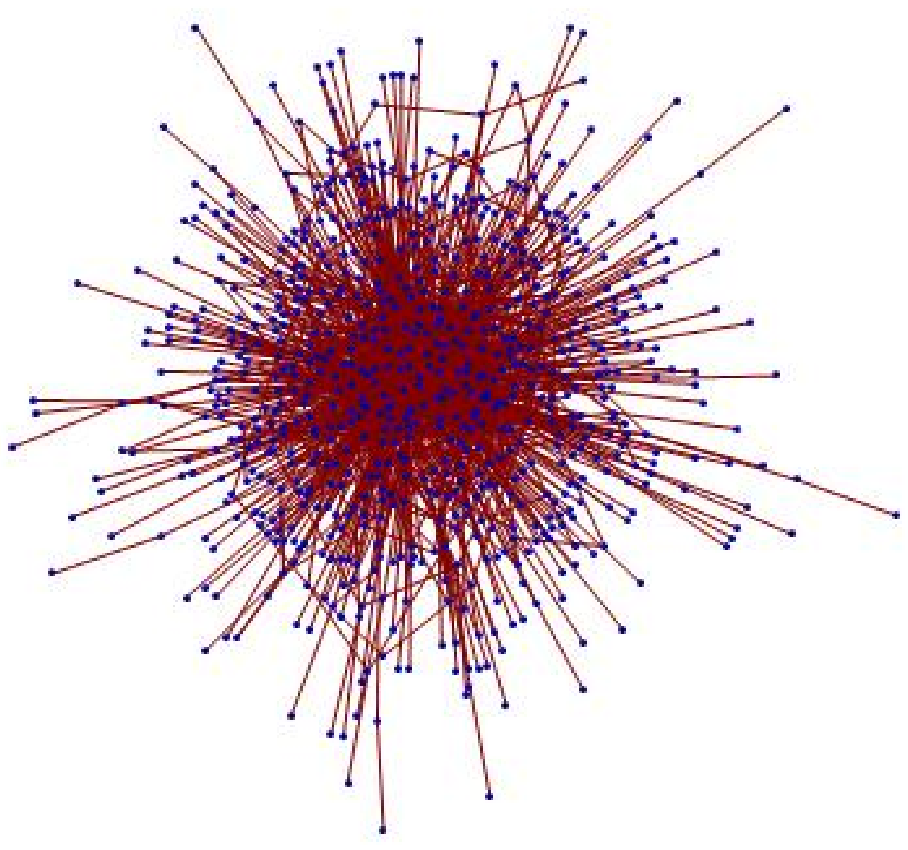}
\end{center}
\caption{Graphical representations of artificially generated
Barab\'asi Albert Scale-free network (741 nodes), S Aureus isB619
(741 nodes) and artificially generated size dependent network (741
nodes).  All figures are generated using the Mathematica spring
embedding algorithm.} \label{fig:NetPics}
\end{figure*}

Although somewhat circumstantial as evidence, it is often useful to
illustrate the similarities and differences between graphical
structures using pictorial representations.   Such comparison
between the original BA model, S. Aureus and the size dependent
growth model demonstrate the similarity between the size dependent
model and the metabolic pathways (see figure \ref{fig:NetPics}).
Here the highly connected initial growth stage, although omitting
the underlying compartmentalisation of the S. Aureus metabolic
network provides a good approximation of the structure. Additionally
the presence of some 'dead end' metabolites, which appear as nodes
connected to only $1$ other node (thus giving rise to higher maximum
path lengths than may otherwise be observed) is not modelled by this
approach.

\section{Discussion}

This investigation has demonstrated that the metabolic networks of
microorganisms are more accurately modelled with a network growth
model in which network size is a modifying factor, a concept which
has surprisingly not previously been considered. This model, as well
as fitting the graphical measures examined, also demonstrates a one
possible mechanism which may be significant in the network
evolution.   This suggests that when the metabolic networks are
growing (or evolving) the size of the network causes it to be
increasingly unlikely for a new metabolite to join the network and
participate in the reactions. Due to the size dependent growth
networks having a densely connected cluster, these networks will
have an increased resilience to targeted attacks than that of the BA
model, which are known to be devastated by targeted attacks
\cite{newman2003structure}.  This suggests that the structure of
metabolic pathways gives them a greater resilience to targeted
attacks than if they were examples of scale-free networks, modelled
by preferential attachment.

The value of the rate at which the probabilities of attachment
decay, $C$, has in our modelling been chosen as a single value to
fit all metabolic networks, however we envisage that for specific
networks a better fit would be available using particular values.

We have presented a model which is both simple and biologically
plausible.  Due to the fact it does not require any specific seed
network it allows for a generic model which can be used to model
various metabolic pathways with only the network size being known.
This allows for various graphical measures to be estimated for any
given metabolic network, without them needing to be individually
analysed.

Clearly the concept of size dependent growth may not be confined to
evolving metabolic networks but may be applicable to a variety of
networks where growth rates may be affected by limited resources.
One example where this may be applicable is that of the London
Underground, where an initially a highly connected network has grown
and now as the network has become larger and more complex it has
become increasingly difficult for new stations and connections to be
added to the network.
\newpage
\pagebreak
\section*{Acknowledgements}

KS is grateful for the financial support of the EU FP7 (KBBE) grant
289434 ``BioPreDyn: New Bioinformatics Methods and Tools for
Data-Driven Predictive Dynamic Modelling in Biotechnological
Applications''.

\bibliographystyle{unsrt}
\bibliography{bib2}

\end{document}